\documentclass[aps,prd,twocolumn,showpacs,nofootinbib]{revtex4}
\usepackage{graphicx}

\begin{document}
\title{Problem of the noise-noise correlation\\
function in hot non-Abelian plasma}
\author{Yu.A. Markov}
\email{markov@icc.ru}
\author{M.A. Markova}
\email{markov@icc.ru}
\affiliation{Institute of System Dynamics\\
and Control Theory Siberian Branch\\
of Academy of Sciences of Russia,\\
P.O. Box 1233, 664033 Irkutsk, Russia}

\date{\today}

\begin{abstract}
In this work on the basis of Kadomtsev's kinetic fluctuation theory
we present the more general expression for noise-noise correlation function
in effective theory for ultrasoft field modes.
\end{abstract}

\maketitle{}

Dynamical processes occurring in systems, described within the
framework of Standard Model at finite temperature (probably with
the minimal supersymmetric extension) play an essential role in
the physics of the early Universe and of heavy ion collisions. In
the weak coupling regime hot non-Abelian gauge theories possess
several energy scale: the hard scale, corresponding to momentum of
order of temperature $T$; the soft scale $\sim gT$ ($g$ is the
gauge coupling) and ultrasoft scale $\sim g^2 T$. As long as we
are interested in the collective excitations with wavelength $\sim
1/gT$, we can ignore, in leading order in $g$, collisions among
the plasma particles \cite{blaizot1}. However the collisions
become a dominant effect for color excitations with wavelength
$\sim 1/g^2 T$.

As known \cite{linde} the color fluctuations characterized by the
momentum scale $g^2T$ are non-perturbative. Their dynamics is of
particular interest, because it is responsible for the large rate
of baryon number violation in hot electroweak theory due to
topology changing transitions of the weak $SU(2)$ gauge fields
\cite{rubakov}. This rate is determined by certain different-time
correlation function of the product of two operators, which in turn are 
a gauge invariant nonlinear functions of the ultrasoft gauge 
fields $A^{a}_{\mu}(X)$. At present the only known instrument to evaluate real 
time dependent quantities is the classical field approximation
\cite{grigoriev} and possible extension \cite{mclerran} which
contains additional degrees of freedom representing the hard field
modes. For time dependent correlation function it was important to find
an effective theory for the ultrasoft field modes.

The effective theory at ultrasoft momentum scale
($\omega\sim g^4T,\;|{\bf p}|\sim g^2T$) is generated by a
Boltzmann-Langevin equation which includes a collision term for color
relaxation and the Gaussian noise term, which keeps the ultrasoft modes
in thermal equilibrium. The Boltzmann-Langevin
equation has been obtained by different approaches. The first is
connected  with B\"odeker's effective theory for $|{\bf p}|\ll gT$ field
modes \cite{bodeker1, bodeker2}. Starting from the collisionless non-Abelian
Vlasov equation, that is the result of integrating out the scale $T$
\cite{blaizot1}, B\"odeker has shown how one can integrate out the scale $g T$
in an expansion in the coupling $g$. At leading order in $g$, he
obtained the linearized Vlasov-Boltzmann equation for the hard field modes,
which besides a collision term also includes a Gaussian noise
arising from thermal fluctuations of initial conditions of the soft fields.
Afterwards, an alternative derivation of the Boltzmann-Langevin equation
was proposed by Litim and Manuel \cite{litim1, litim2}. The authors used
a classical transport theory in the spirit of Heinz \cite{heinz}.
The approach of Litim and Manuel
\cite{litim1, litim2} provides not only the correct collision term
but also the correct noise-noise correlator. This correlator was
obtained, similar to B\"odeker \cite{bodeker1}, directly from
the microscopic theory without making use of the
fluctuation-dissipation theorem. A somewhat different
approach of more phenomenological character to the computation of the
correlator of stochastic source was presented by the same authors
in \cite{litim3}, where the well known link between a linearized
collision integral and the entropy was exploited.

Blaizot and Iancu \cite{blaizot2} presented a detailed derivation of the
Vlasov-Boltzmann equation, starting from the Kadanoff-Baym equations. The
derivation is based on the method of gauge covariant gradient expansion
first proposed by them for the collective dynamics at the scale $g T$
\cite{blaizot1}. In work \cite{blaizot3} Blaizot and Iancu
derived the statistics of the noise term in the Boltzmann-Langevin equation
by using the fluctuation-dissipation theorem together with the known
structure of the collision term in the Boltzmann equation.

The purpose of this paper is to show that the Boltzmann-Langevin equation
in the form obtained by Blaizot and Iancu  \cite{blaizot2}
is somewhat more general, than the equation obtained by B\"odeker
\cite{bodeker1, bodeker2}.

We use the metric $g^{\mu \nu} = diag(1,-1,-1,-1)$, choose units such
that $c=k_{B}=1$ and note $X=(X_0,{\bf X}).$
On a space-time scale $X\gg (gT)^{-1}$ the ultrasoft colored fluctuation of the
gluon density in the adjoint representation
$\delta N({\bf k},X)=\delta N^a({\bf k},X)T^a$ $((T^a)^{bc}\equiv -if^{abc})$
satisfies the linearized
Boltzmann-Langevin equation\footnote{This equation is taken in the form
suggested by Blaizot and Iancu in Ref.\,\cite{blaizot2}.}
\begin{equation}
[v \cdot D_X, \delta N({\bf k}, X)] + g{\bf v}\cdot {\bf E}(X)
\,\frac{dN(\epsilon_{\bf k})}{d\epsilon_{\bf k}}
\label{eq:q}
\end{equation}
\[
=
\hat{\rm C}_{\bf k}\delta N({\bf k},X) + y({\bf k},X).
\]
Here, $v=(1,{\bf v}),\;{\bf v}={\bf k}/\vert{\bf k}\vert$;
$D_{\mu} = \partial_{\mu} + igA_{\mu}(X)$;
$[\,,\,]$ denotes a commutator; ${\bf k}$ is a momentum of hard thermal gluons;
${\bf E} (X) = {\bf E}^a(X) T^a$ is a chromoelectric field;
$N(\epsilon_{\bf k}) = 1/(\exp(\epsilon_{\bf k}/T) - 1)$ is a boson
occupation factor, where $\epsilon_{\bf k}\equiv\vert{\bf k}\vert$.
The collision operator $\hat{\rm C}_{\bf k}$ acts on function on the right
according to \cite{blaizot2}
\begin{equation}
\hat{\rm C}_{\bf k}f({\bf k})\equiv
g^4N_cT\!\int\!\frac{d {\bf k}^{\prime}}{(2\pi)^3}\,
\Phi ({\bf v}\cdot{\bf v}^{\prime})
\label{eq:w}
\end{equation}
\[
\times
\biggl\{\frac{dN(\epsilon_{{\bf k}^{\prime}})}
{d\epsilon_{{\bf k}^{\prime}}}
\,[T^a,[T^a,f({\bf k})]]-
\frac{dN(\epsilon_{\bf k})}{d\epsilon_{\bf k}}\,T^a\,{\rm Tr}\,
(T^af({\bf k}^{\prime}))\biggr\},
\]
where the collision kernel $\Phi ({\bf v}\cdot{\bf v}^{\prime})$ reads
\[
\Phi ({\bf v}\cdot{\bf v}^{\prime})\!\simeq\!
\frac{2}{\pi^2m_D^2}
\frac{({\bf v}\cdot{\bf v}^{\prime})^2}
{\sqrt{1- ({\bf v}\cdot{\bf v}^{\prime})^2}}\ln\left(\frac{1}{g}\right)\!,
\,m_D^2= \frac{1}{3}\,g^2N_cT^2
\]
within logarithmic accuracy. The function $y({\bf k},X)=y^a({\bf k},X)T^a$
on the right-hand side of Eq.\,(\ref{eq:q}) is a noise term.
This term injects energy compensating the energy loss at scale $g^2T$ by virtue
of the damping term.

Furthermore we write out a general expression for a correlation
function of the noise term $y({\bf k},X)$ in the form proposed by
Kadomtsev \cite{kadom} with a minimal extension to the color degrees
of freedom
\[
\ll\!y^a({\bf k}, X) T^a\!\otimes y^b({\bf k}^{\prime},X^{\prime})T^b\!\gg\,
\]
\begin{equation}
=-\frac{1}{2N_c}\,
\Bigl(\hat{\rm C}_{\bf k}\otimes \hat{\rm I} +
\hat{\rm I}\otimes\hat{\rm C}_{{\bf k}^{\prime}}\Bigr)T^a\!\otimes T^a
\label{eq:e}
\end{equation}
\[
\times\,(2\pi)^3\delta^{(3)}\!({\bf k}-{\bf k}^{\prime})
N(\epsilon_{{\bf k}^{\prime}})[1+N(\epsilon_{{\bf k}^{\prime}})]
\,\delta^{(4)}\!(X-X^{\prime}).
\]
Here, a symbol $\otimes$ denotes a direct production in a color space,
$\hat{\rm I}$ is an identity operator. We note that in original work of
Kadomtsev \cite{kadom} the factor $N(\epsilon_{{\bf k}^{\prime}})$ in
noise-noise correlation function stands instead
of the factor $N(\epsilon_{{\bf k}^{\prime}})
[1+N(\epsilon_{{\bf k}^{\prime}})]=
-T(dN(\epsilon_{{\bf k}^{\prime}})/d\epsilon_{{\bf k}^{\prime}})$.
In \cite{kadom} pure classical gas with
Maxwell-Boltzmann statistic was considered, while we consider a hot
quantum plasma for gluons (in the semiclassical limit), which obey
Bose-Einstein statistic.
Using the definition of collision term (\ref{eq:w}) and decomposing the
momentum $\delta$-function in polar coordinates
$$
\delta^{(3)}\!({\bf k}-{\bf k}^{\prime})=
\frac{1}{4\pi{\bf k}^2}\;\delta(|{\bf k}|-|{\bf k}^{\prime}|)\,
\delta^{(S^2)}\!({\bf v}-{\bf v}^{\prime}),
$$
where $\delta^{(S^2)}\!({\bf v}-{\bf v}^{\prime})$ is a delta-function on a
unit sphere,
from (\ref{eq:e}) we obtain the following expression for the noise-noise
correlation function
\begin{widetext}
\begin{equation}
\ll\!y^a({\bf k}, X) y^b({\bf k}^{\prime},X^{\prime})\!\gg
= -\,(2\pi)^3 \delta^{ab}\frac{T}{N_c}\,
\Biggl\{\gamma\,\frac{dN(\epsilon_{\bf k})}{d\epsilon_{\bf k}}\,
\frac{1}{4\pi{\bf k}^2}\;\delta(|{\bf k}|-|{\bf k}^{\prime}|)\,
\delta^{(S^2)}\!({\bf v}-{\bf v}^{\prime})
+\, \frac{g^4N_c^2T}{(2\pi)^3}\;
\frac{dN(\epsilon_{\bf k})}{d\epsilon_{\bf k}}\,
\frac{dN(\epsilon_{{\bf k}^{\prime}})}{d\epsilon_{{\bf k}^{\prime}}}\;
\Phi ({\bf v}\cdot{\bf v}^{\prime})\Biggr\}\,
\]
\[
\times\,\delta^{(4)}\!(X-X^{\prime}).
\label{eq:r}
\end{equation}
\end{widetext}
Here,
\[
\gamma\!=\!m_D^2\,\frac{g^2N_cT}{2}\!\!
\int\!\!\frac{d\Omega_{{\bf v}^{\prime}}}{4\pi}\,
\Phi ({\bf v}\cdot{\bf v}^{\prime})
\!\simeq\!
\frac{g^2N_cT}{2}\,\biggl(\!\ln\Bigl(\frac{m_D}{\mu}\Bigr)+O(1)\!\biggr)
\]
is the damping rate for a hard transverse gluon with velocity
${\bf v}$, and $\mu$ is the magnetic screening ``mass'', usually entered
by hand for removal of infrared divergence. The equation (4) is
the main result of our report.

From Eq.\,(4) we see, that the noise term $y({\bf k},X)$
depends on both the velocity ${\bf v}$ (unit vector) and the
magnitude $|{\bf k}|$ of the momentum in a nontrivial way, and
thus generates (by virtue of the Boltzmann-Langevin equation
(\ref{eq:q})) similar dependence for the fluctuation $\delta N({\bf
k},X)$. Note that Litim and Manuel in \cite{litim3} (Eqs.\,(23)
and (32)) pointed to the possible nontrivial dependence of
noise-noise correlator on the magnitude $|\textbf{k}|$ of the
momentum. However our expression (4) is more complicated,
since here contrary to \cite{litim3}, we have different factors
(depending on $|\bf{k}|$ and $|\bf{k^{\prime}}|$) before functions
$\delta^{(S^2)}\!({\bf v}-{\bf v}^{\prime})$ and $\Phi ({\bf
v}\cdot{\bf v}^{\prime})$.

For the calculation of the color current
\[
j_{\mu}(X) = 2gN_c\!\int\!\!\frac{d{\bf k}}{(2 \pi)^3}
\,v_{\mu}\delta N({\bf k},X)
\]
we need only the second momentum with respect to the
magnitude $|{\bf k}|$ of ultrasoft fluctuation $\delta N({\bf k},X)$.
For this and similar physical problems, where the second
moment is the only relevant quantity, it is convenient to
introduce new functions  $W(X,{\bf v})$ and $\nu(X,{\bf v})$
depending on velocity ${\bf v}$ only,
instead of initial functions $\delta N({\bf k},X)$ and $y({\bf k},X)$,
by the rules
\begin{equation}
\int\limits_0^{\infty}\!{\bf k}^2d|{\bf k}|\,\delta N({\bf k},X)
=-g\,W(X,{\bf v})\!
\int\limits_0^{\infty}\!{\bf k}^2d|{\bf k}|
\,\frac{dN(\epsilon_{\bf k})}{d\epsilon_{\bf k}},
\label{eq:t}
\end{equation}
\begin{equation}
\int\limits_0^{\infty}\!{\bf k}^2d|{\bf k}|\,y({\bf k},X) =
-g\,\nu(X,{\bf v})\!
\int\limits_0^{\infty}\!{\bf k}^2d|{\bf k}|\,
\frac{dN(\epsilon_{\bf k})}{d\epsilon_{\bf k}}.
\hspace{0.5cm}
\label{eq:y}
\end{equation}
Here, the first relation is an extension of usually used parametrization of
off-equilibrium fluctuation \cite{baym, blaizot2}
\[
\delta N({\bf k},X) = -g\,W(X,{\bf v})\,
\frac{dN(\epsilon_{\bf k})}{d\epsilon_{\bf k}},
\]
which is valid in the absence of the noise term $y({\bf k},X)$ or
when this term depends on velocity ${\bf v}$ only. The general
connection, Eq.\,(\ref{eq:t}), between the functions $\delta N({\bf k},X)$ and
$W(X,{\bf v})$ involves an integral over ${\bf k}$, which reflects the
corresponding integral in the relation (\ref{eq:y}) between the noise term
$y({\bf k},X)$ and $\nu(X,{\bf v})$.

Multiplying Eq.\,(\ref{eq:q}) and correlation function (4) by
${\bf k}^2$ and ${\bf k}^2 {{\bf k}^{\prime}}^2$, and
integrating over $d|{\bf k}|$ and $d|{\bf k}| d|{\bf k}^{\prime}|$,
correspondingly (with regard to (\ref{eq:t}) and (\ref{eq:y})) instead
of Eq.\,(\ref{eq:q})\,--\,(\ref{eq:e}) we recover the equations
for the functions $W(X,{\bf v})$ and $\nu(X,{\bf v})$,
first proposed by B\"odeker \cite{bodeker1, bodeker2}.
Let us stress that such a reduction of initial system of
Eqs.\,(\ref{eq:q})\,--\,(\ref{eq:e}) to simpler system
for functions $W(X,{\bf v})$ and $\nu(X,{\bf v})$ does not lead to any loss
of the information, if we only restrict our consideration to the second
moment with respect to $|{\bf k}|$ of ultrasoft gluon fluctuations
$\delta N({\bf k},X)$.
But for the calculation of more general quantities, which involve also the
other moments of the ultrasoft fluctuations (e.g. correlation function of
energy density fluctuation), there is no such one-to-one correspondence
between these systems by virtue of nontrivial dependence of noise-noise
correlator (4) on magnitudes $|{\bf k}|$ and $|{\bf k}^{\prime}|$.
In this case it is necessary to use more exact system of equations (\ref{eq:q}),
(\ref{eq:w}) and (4).

\section*{\bf Acknowledgments}
This work was supported by the Russian Foundation for Basic Research
(project no 03-02-16797).

\end{document}